\documentclass[aip,amsmath,amssymb,reprint]{revtex4-1}

\usepackage{graphicx}
\usepackage{dcolumn}
\usepackage{bm}

\usepackage[utf8]{inputenc}
\usepackage[T1]{fontenc}
\usepackage{mathptmx}
\usepackage{bm}
\usepackage{bbm}
\usepackage{amsfonts}
\usepackage{amsmath}
\usepackage{natbib}
\usepackage{dsfont} 
\providecommand{\keywords}[1]{\textbf{\textit{Index terms---}} #1}
\usepackage[mathlines]{lineno}
\usepackage[colorinlistoftodos]{todonotes}
\usepackage[colorlinks=true, allcolors=blue]{hyperref}
\usepackage{soul,xcolor}
\usepackage{siunitx}
\usepackage{braket}
\usepackage{textcomp}
\setstcolor{red}
\RequirePackage{color}

\newcommand{\VI}[1]{{\color{blue}{#1}}}

\begin{document}

\title[Ballistic guided electrons against disorder in graphene nanoribbons]{Ballistic guided electrons against disorder in graphene nanoribbons}
\author{E. J. Robles-Raygoza}
 \affiliation{Facultad de Ciencias, Universidad Autónoma de Baja California, Apartado postal 1880, 22800 Ensenada, Baja California, México}%
\author{V. G. Ibarra-Sierra}%
\email{ibarrasierra@uabc.edu.mx}
\affiliation{Facultad de Ciencias, Universidad Autónoma de Baja California, Apartado postal 1880, 22800 Ensenada, Baja California, México}%
\author{J. C. Sandoval-Santana}%
\affiliation{Centro de Nanociencias y Nanotecnolog\'ia, Universidad Nacional Aut\'onoma de M\'exico, Apartado Postal 2681, 22800 Ensenada, Baja California, M\'exico.}%
\author{R. Carrillo-Bastos}
\affiliation{Facultad de Ciencias, Universidad Autónoma de Baja California, Apartado postal 1880, 22800 Ensenada, Baja California, México}%

\preprint{AIP/123-QED}

\date{\today}

\begin{abstract}
Graphene nanoribbons (GNRs) are natural waveguides for electrons in graphene. Nevertheless, unlike micron-sized samples, conductance is nearly suppressed in these narrow graphene stripes, mainly due to scattering with edge disorder generated during synthesis or cut. A possible way to circumvent this effect is to define an internal waveguide that isolates specific modes from the edge disorder and allows ballistic conductance. There are several proposals for defining waveguides in graphene; in this manuscript, we consider strain folds and scalar potentials and numerically evaluate these proposals' performance against edge and bulk disorder. Using the Green's function approach, we calculate conductance and the local density of states (LDOS) of zigzag GNRs and characterize the performance of these different physical waveguiding effects in both types of disorder. We found a general improvement in the electronic conductance of GNR due to the presence of the internal waveguiding, with the emergence of plateaus with quasi-ballistic properties and robustness against edge disorder. These findings are up to be applied in modern nanotechnology and being experimentally tested.
\end{abstract}

\maketitle

\section{Introduction} \label{sec:introduction}

The electronic, optical, chemical and mechanical properties of graphene\cite{Stauber2007,Balandin2008,Neto2009,Abergel2010,sarma2011electronic,Sturala2018} continue to offer new possibilities for developing technological applications\cite{Wonbong2011,Randviir2014}. In particular, the theoretical and experimental analysis of graphene nanoribbons (GNRs) is of great significance due to its electronic properties and potential use in the design of electronic devices\cite{Mucciolo2009,Dutta2010,wakabayashi2010electronic,torres2014introduction,bischoff2015localized,carrillo2016strained,Celis_2016,Peeters2016,Tejada2016,kim2016valley}. Likewise, several studies have shown that quantum ballistic transport at low temperatures strongly depends on edge topology or boundary conditions on GNRs, namely the armchair or zigzag edges\cite{Fujita1996,Nakada1996EdgeState,torres2014introduction}. Usually, edge and bulk imperfections from the experimental synthesis of these GNRs drastically suppress electronic conductance and ballistic transport\cite{munoz2006coherent,Areshkin2007,Gunlycke2007emiconducting,lherbier2008transport,Evaldsson2008,Li2008QuantumConductance,Busu2008Effect,martin2009transport,Mucciolo2009,mucciolo2010disorder,bischoff2015localized}. Such imperfections can be theoretically modeled through vacancies or random scalar potentials.  For example, in the case of edge disorder, the usual approach is based on randomly removing successive sites on GNR edges\cite{Li2008QuantumConductance,Busu2008Effect,Cresti_2009,Mucciolo2009}. Concerning bulk disorder, typical numerical studies introduce Gaussian potentials with random amplitudes and uniform distribution\cite{Lewenkopf2008,Mucciolo2009,Wurm2012}. Experimental confirmation of this models can be obtained with electronic transport measurements on GNRs with cesium atoms deposited on top as the source of bulk disorder \cite{Smith2013}. 

Scalar potentials and strain deformations are possible ways to control the electronic transport in graphene\cite{Querlioz2008Suppression,ribeiro2009strained,lu2010band,Naumis_2017}; in particular, they allow the definition of an internal waveguide in graphene nanoribbons to avoid the scattering by edge disorder\cite{ChengPRL2019}. For example, uniaxial gaussian strain folds generate new transport channels where electrons travel through the deformed region showing robustness against edge disorder\cite{Carrillo2014Gaussian,carrillo2016strained}. More recent research opens up the possibility of tuning the properties in graphene/hBN  through the deformation of flakes and wrinkles\cite{Giambastiani2022}. Other schemes of implementing waveguides are local gate voltages\cite{Hartmann2010Waveguides,Cao2017Waveguides,Mosallanejad2018Waveguides} or using positive and negative charge nanotubes located at a certain distance on the graphene sample\cite{Hartmann2020Waveguides}; in any case, the number of accessible modes depends on the potential's intensity, width, and shape\cite{Hartmann2010Waveguides,Hartmann2014Waveguides,Ying2015Waveguide,Rickhaus2015,Cao2017Waveguides,Mosallanejad2018Waveguides,Zubair2019Waveguides,Shah2019Waveguides,Zhang2019Waveguides,Hartmann2020Waveguides}. The basic principle of waveguides is to create confined bound states in the transversal direction (localized modes) while keeping the dispersive nature in the other, akin to the transport of photons in optical fiber\cite{Friebele1985,Addanki2018Review,Dragic2018}. 

In this work, we study the conductance behavior in zigzag graphene nanoribbons (zGNRs) in the presence of an internal waveguide generated by a Gaussian strain fold or local scalar potentials and considering the effect of edge and bulk disorder\cite{Mucciolo2009,mucciolo2010disorder}. Following the standard recursive Green function approach based on the tight-binding description of the system, we obtain the two-terminal conductance and the local density of states (a similar approach to Refs. [\onlinecite{Mucciolo2009}] and [\onlinecite{Lewenkopf2013}]). We show that the internal waveguiding generally changes the energy at which certain plateaus appear, modifying the number of available channels at a given energy\cite{carrillo2016strained}. Some of these new available channels travels within the internal waveguide avoiding the scattering with edge disorder. Therefore, although in the presence of edge disorder the conductance at the first plateau almost vanishes because the edge states are strongly affected;  plateaus associated with the internal waveguide show general robustness against edge disorder. We compare the performance of these different waveguiding effects against edge and bulk disorder. We consider a strain defined waveguide\cite{Carrillo2014Gaussian} and three different electrostatic waveguides: a square potential well\cite{cao2017investigation}, hyperbolic-secant potential well\cite{Hartmann2010Waveguides}, and bipolar\cite{Hartmann2020Waveguides}  (composed of attractive and repulsive hyperbolic potentials), as shown schematically in Fig. \ref{fig1:Nanoribbon}(a).

The paper is organized as follows. In Sec. \ref{sec:II}, we introduce general model to study the quantum conductance on zGNRs in presence of waveguides, edge disorder and bulk disorder. Subsequently, in Sec. \ref{sec:III} we study the conductance using a strain Gaussian fold, square potential well, hyperbolic-secant potential well and bipolar potential and only in presence of edge disorder. In Sec. \ref{sec:IV} we include the joint effect of edge and bulk disorder in the studio of waveguides. Finally, we summarize and conclude in
Sec.\ref{sec:V}.
\section{General Model} \label{sec:II}

\begin{figure}[t]
\includegraphics[width=.47\textwidth]{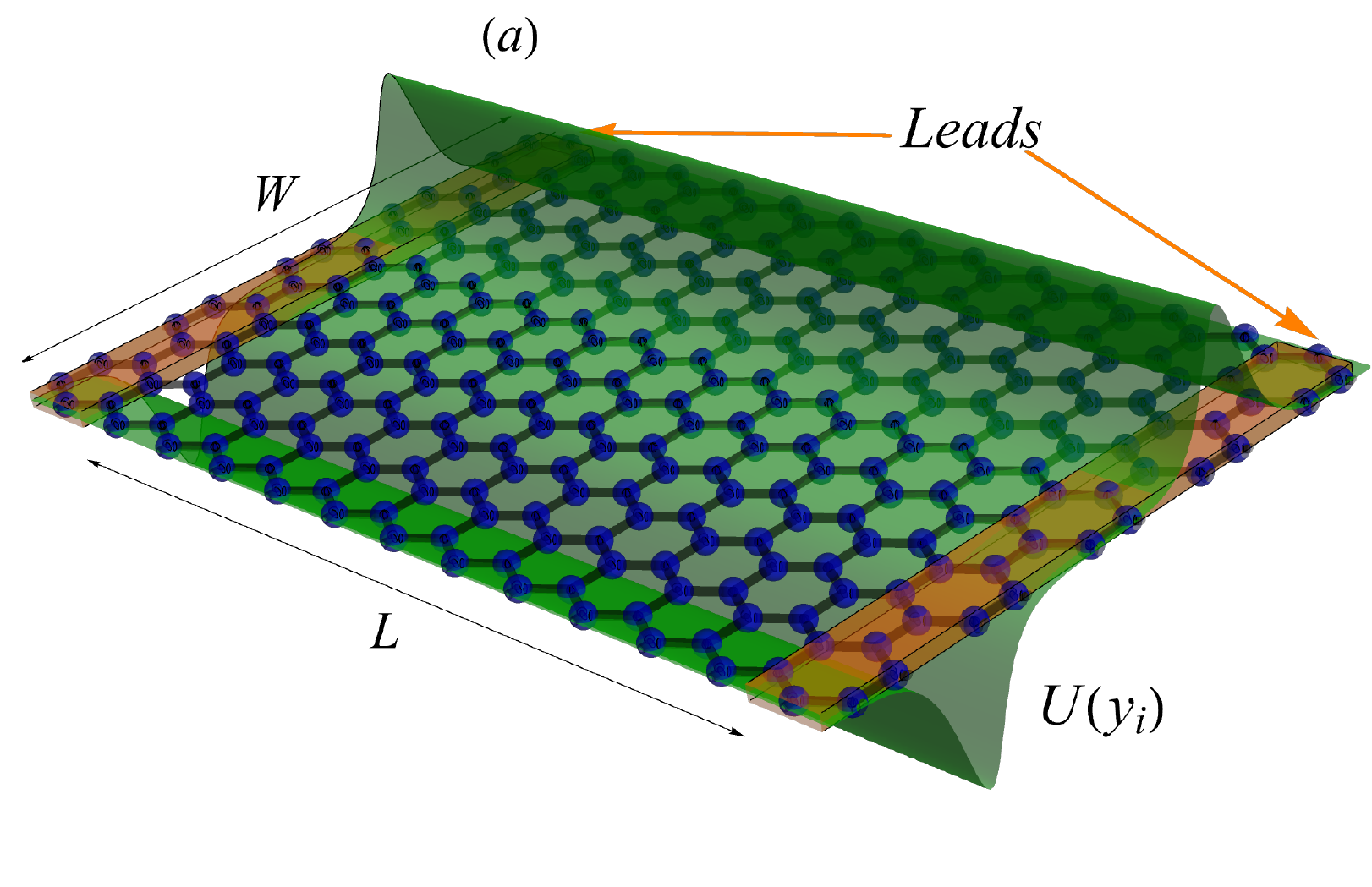}
\includegraphics[width=.47\textwidth]{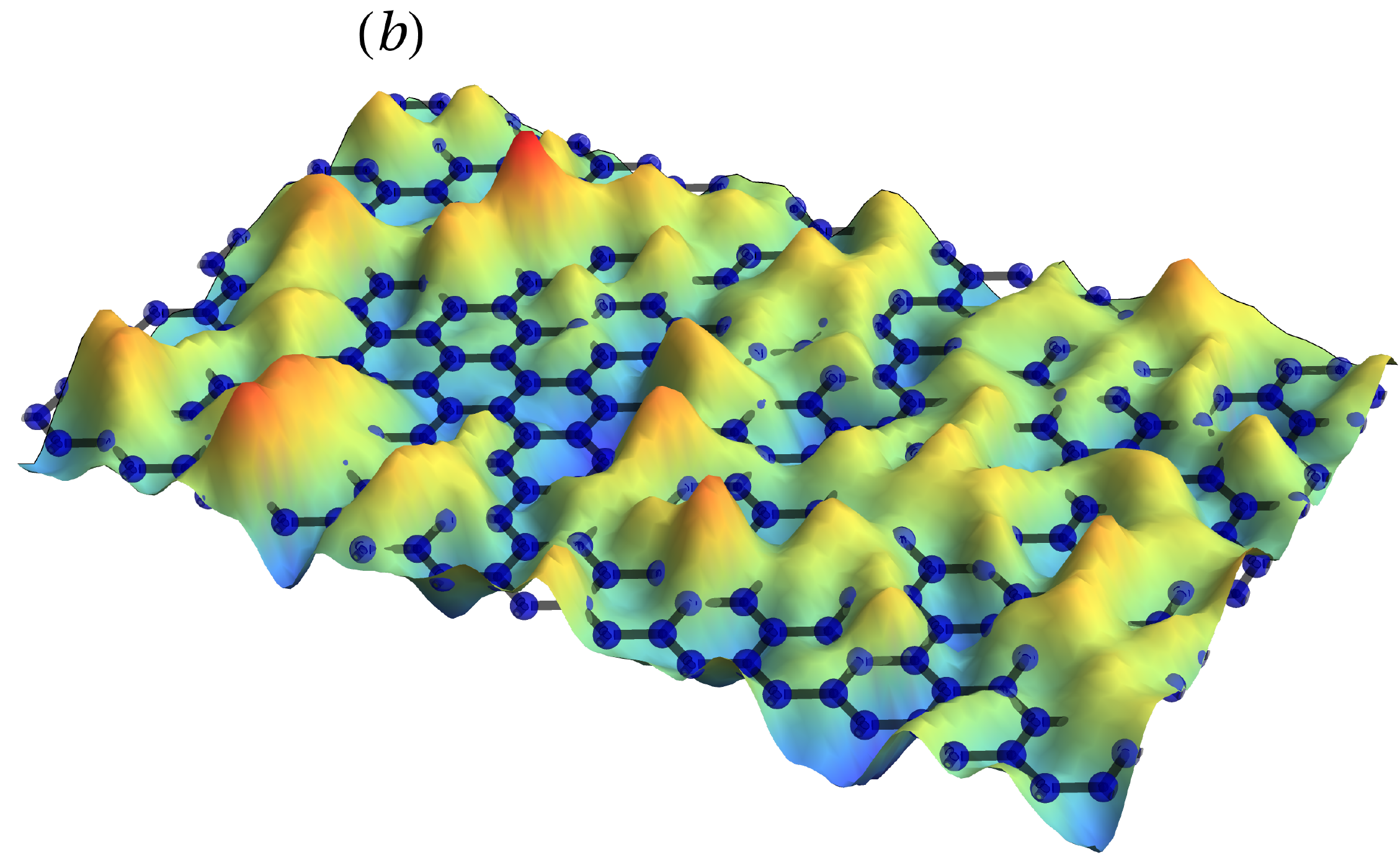}
\includegraphics[width=.47\textwidth]{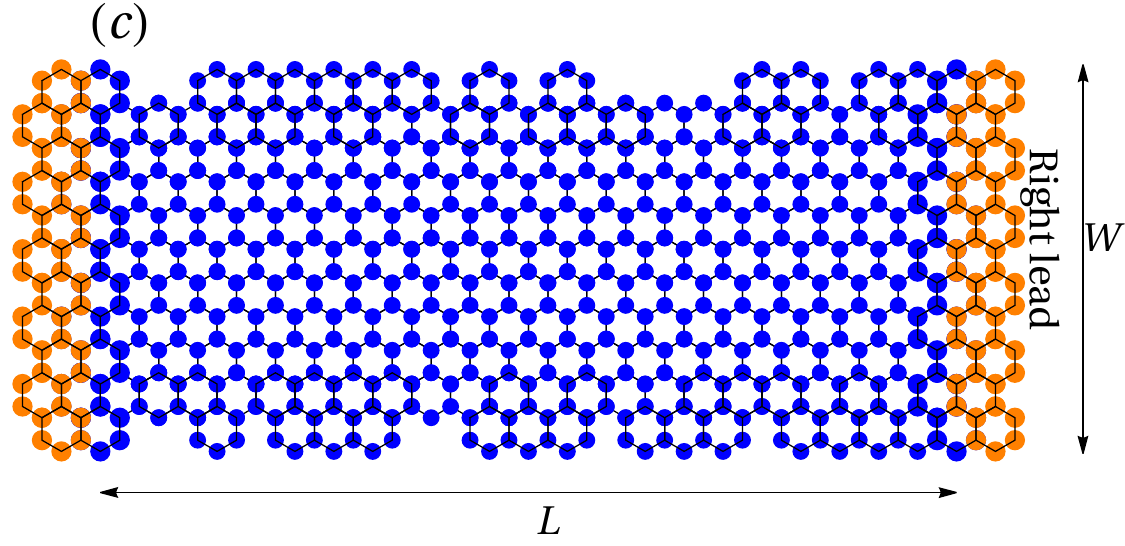}
\caption{\label{fig1:Nanoribbon} (Color) Schematic diagrams of zGNRs with length $L$ and width $W$ connected to two leads. The nanoribbon is subject to (a) general potential $U(y_i)$ describes by Eq. (\ref{eq:Potentials}), (b) under a potential $U_{\text{imp}}(\mathbf{r_i})$ of bulk disorder and (c) varying edge roughness.}
\end{figure}

The system is a zGNR with length $L$ and width $W$ connected to graphene leads, as schematically shown in Fig \ref{fig1:Nanoribbon}(a). We evaluate the mechanisms that generate the waveguides to establish which one conserves the ballistic transport in the presence of edge and bulk disorders.  

The electron dynamics in the zGNR is described by the tight-binding Hamiltonian\cite{bena2009remarks,wakabayashi2010electronic}
\begin{equation}\label{eqn:Hamiltonian}
H=\sum_{\langle i,j \rangle}\left[U_{i} c_{i}^{\dagger}c_{i}+t_{ij}c_{i}^{\dagger}c_{j}\right]+h.c.,
\end{equation}
where $U_i=U\left(y_{i}\right)+U_{\text{imp}}(\mathbf{r_i})$ is the modified site energy at the $i$-th site due to an electrostatic waveguide potential $[U\left(y_{i}\right)]$ and  the on-site disorder potential $[U_{\text{imp}}(\mathbf{r_i})]$ in the zGNR. 

We model the waveguide potential $U\left(y_{i}\right)$ as
\begin{equation}\label{eq:Potentials}
U\left(y_{i}\right)=-u_1 \mathrm{sech}\left[\frac{\left(y_{i}+d_1\right)}{W_1}\right]+ u_2 \mathrm{sech}\left[\frac{\left(y_{i}-d_2\right)}{W_1}\right]+f(y_i), \end{equation}
with
\begin{equation}\label{eqn:Perfil_Poo}
f(y_i)=\left\{
\begin{array}{lll}
0     &  {\rm if } & -W/2\leq y_i\leq -W_2/2, \\
-u_3   & {\rm if} &  -W_2/2 < y_i < W_2/2, \\
0     &   {\rm if} &  W_2/2 \leq y_i\leq W/2.
\end{array}
\right.
\end{equation}
The first two elements in Eq. \eqref{eq:Potentials} describe an attractive and repulsive hyperbolic-secant potential, and the third element is the square potential well where $u_{1,2,3}$ are the potential amplitudes, $W_{1,2}$ are the potential widths and $d_{1,2}$ are the potential displacements. A schematic diagram of the potential $U(y_i)$ is shown in Fig.\ref{fig1:Nanoribbon}(a). Bulk disorder is modeled $U_{\text{imp}}(\mathbf{r_i})$  by a random array of Gaussian potential centers with random strength, as described  by\cite{Lewenkopf2008,Rycerz2007,Mucciolo2009,shon1998}
\begin{equation}
U_{\text{imp}}(\mathbf{r_i})=\displaystyle\sum_{n=1}^{N_{\text{imp}}} U_n \exp\left[-\frac{|\mathbf{r_i}-\mathbf{r}_{\text{imp}}|^2}{2 \xi^2}\right],
\end{equation}
where $U_{\text{imp}}(\mathbf{r_i})$ is the sum contributions of impurities for each carbon site and $N_{\text{imp}}$ is the total number of impurities in the zGNRs. The density of impurities is $n_{\text{imp}}=N_{\text{imp}}/N=0.04 \,(4\%)$ with $N$ as the total number atomic sites in the nanoribbon. Similarly, $\mathbf{r_i}$ and $\mathbf{r}_{\text{imp}}$ are the positions of every site and impurity,  respectively. $\xi=2 a$ is the range of disorder potential and $U_n$ is the amplitude of the potential which is randomly distributed within the range $\left[-\delta U,\delta U \right]$ with $\delta U < |(0.1) t_0|$. The schematic diagram of Gaussian bulk disorder for a particular realization is shown in Fig. \ref{fig1:Nanoribbon}(b).

We implement edge disorder as reported in Ref. [\onlinecite{Mucciolo2009}]. This approach consists of randomly removing up to $N_{\text{sweep}}$-th atoms at the edges along the zigzag nanoribbon, as is shown in Fig. \ref{fig1:Nanoribbon}(c). The control parameters in this implementation are the $N_{\text{sweep}}$ which is related to how rough the edges are, and the probability \textit{$p_{k}$} of removing the edge atom in the \textit{k}-th sweep \cite{Mucciolo2009}. In this work, we take $N_{\text{sweep}}=3$ and the probabilities of removing carbon atoms in the edge of zGNR are $p_1=30\%$, $p_2=20\%$ and $p_3=10\%$.

Similarly in Eq. (\ref{eqn:Hamiltonian}), $t_{ij}$ is the hopping energy between the first neighboring sites $i$ and $j$ modified by the presence of an out-of-plane strain deformation.
Hence, we introduce strain as renormalization of the hopping parameter, given by $t_{ij}=t_{0}\exp\left[{-\beta\left(l_{ij}/a-1\right)}\right]$, where $t_{0}=-2.8\,$eV is the hopping parameter in the absence of strain, $a=0.142\,\,$nm is the interatomic distance in unstrained graphene, and $\beta=\mid\partial\log t_{0}/\partial \log a\mid \simeq 3$ is the Gr\"uneisen parameter for graphene \cite{Carrillo2014Gaussian}. The modified interatomic distances are given by
\begin{equation}\label{eq:InteratomicDistance}
l_{ij}=a+\frac{\varepsilon_{yy}}{a}y_{ij}^{2},    
\end{equation}
where $\varepsilon_{yy}=\partial_y h(y)\partial_y h(y)$ is the strain tensor and $y_{ij}$ is the projected distance between the sites $i$ and $j$ in $y$ direction\cite{carrillo2016strained}. The definition of $\varepsilon_{yy}$ includes the out-of-plane deformation function $h(y)=A\exp\left[-(y-y_0)^{2}/b^{2}\right]$ where its center, amplitude and width are given by $y_0=W/2$, $A$, $b$ respectively. 

Finally, the conductance $G(\tilde E)$ and LDOS in zGNRs are calculated using standard recursive Green’s function techniques in graphene connected to leads \cite{Lewenkopf2013,Carrillo2014Gaussian,carrillo2016strained}.

\section{Conductance and edge disorder} \label{sec:III}
\begin{figure}[t]
        \centering
        \includegraphics[width=0.485\textwidth]{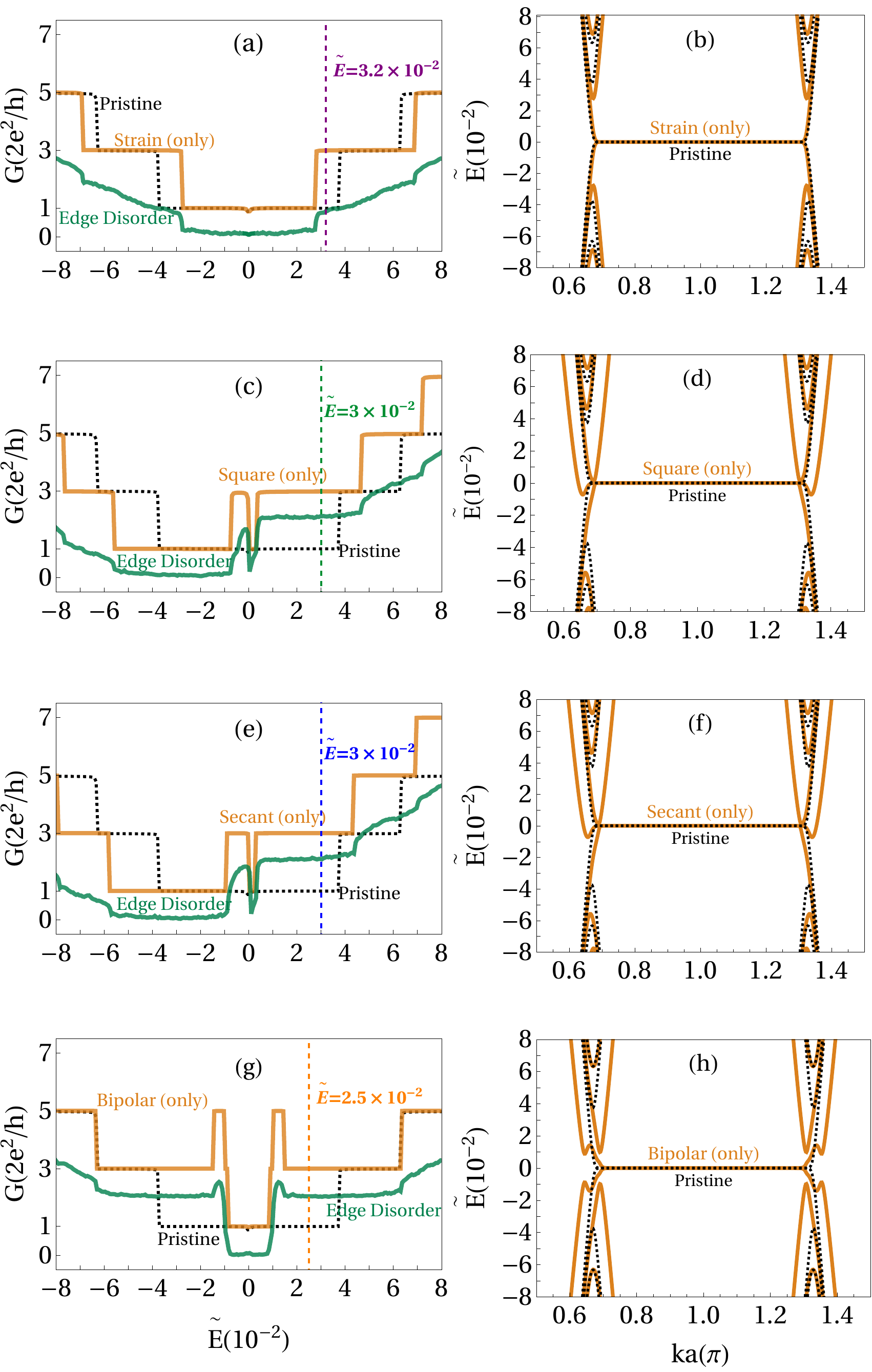}
         \caption{\label{fig:Conductance} (Color) Two terminal conductance ($G(\tilde E)$) and band structure ($\tilde{E}=E/|t_0|$) in a graphene nanoribbon ($L=54.8\,$nm and $W=25.8\,$nm). From the upper to the lower panel, the solid orange line represents the conductance and band structure for (a)-(b) strained, (c)-(d) centered square potential well, (e)-(f) centered hyperbolic-secant potential well, and (g)-(h) bipolar potential. In all left panels, the solid green line describes the additional effect of edge disorder over $100$ realizations in the conductance. Similarly, the dotted black line in all left (right) panels corresponds to the pristine zGNR conductance (band structure). The vertical color dashed lines in (a), (c), (e) and (g) correspond to values of $\tilde{E}$ in Figs. \ref{fig:LDOS}, \ref{fig:GvsL_Potential_EdgeDisoder} and \ref{fig:GvsL_Potential_EdgeBulkDisoder}.}
    \end{figure}
     \begin{figure}[t]
        \centering
        \includegraphics[width=0.485\textwidth]{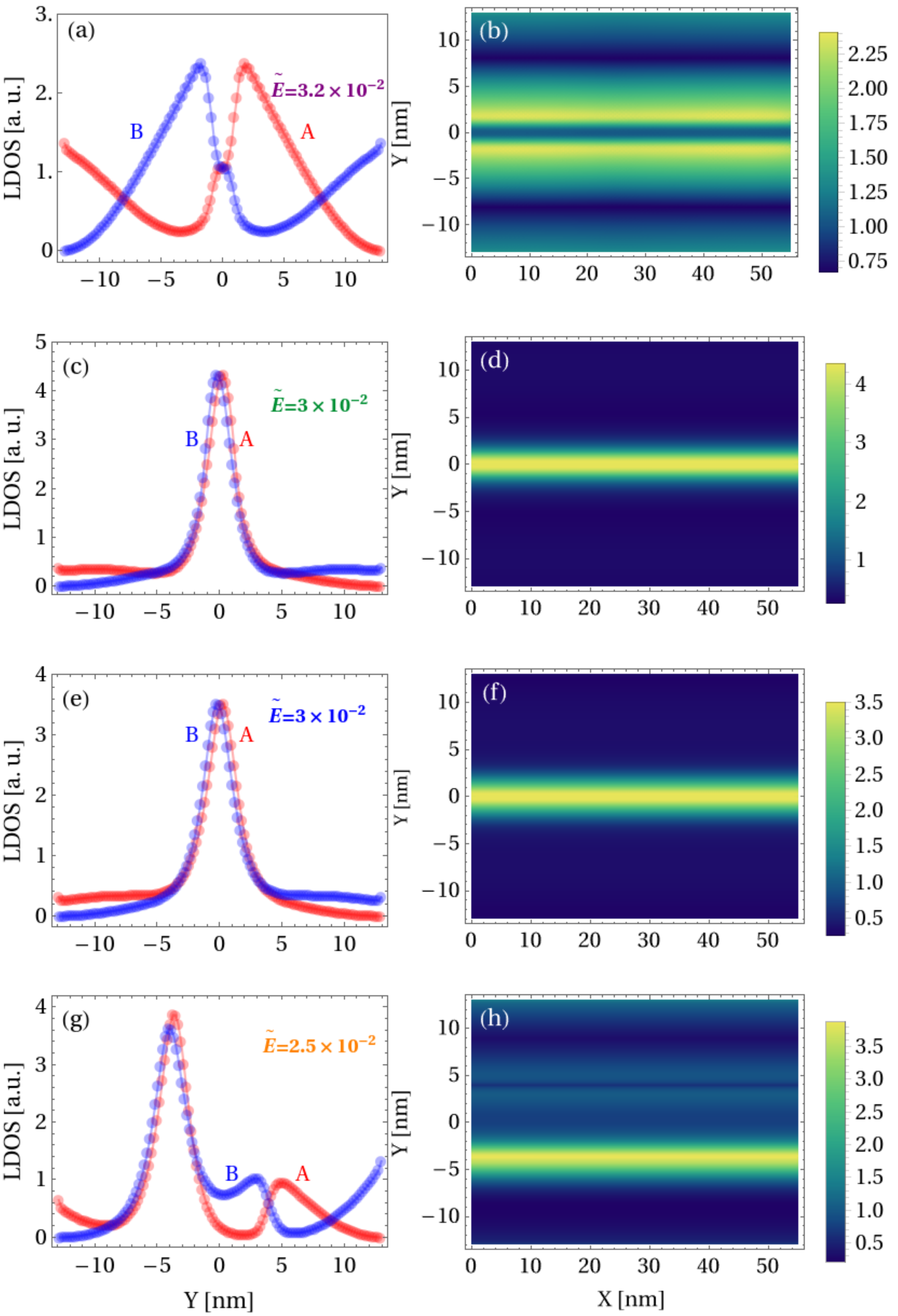}
        \caption{\label{fig:LDOS} (Color) LDOS profile across the nanoribbon (left panels) and LDOS bidimensional color map (right panels). These LDOS are numerically calculated using specific waveguide. In these panels, we show the cases of:  (a)-(b) strain to an energy $\tilde{E}=3.2\times10^{-2}$, (c)-(d) square potential well with $\tilde{E}=3.0\times10^{-2}$, (e)-(f) hyperbolic-secant potential well with $\tilde{E}=3.0\times10^{-2}$ and (g)-(h) bipolar potential with $\tilde{E}=2.5\times10^{-2}$. In all right panels the red (blue) curve correspond to LDOS for sublattice A (B) in zGNRs.}
    \end{figure}
     \begin{figure}[t]
        \centering
        \includegraphics[width=.47\textwidth]{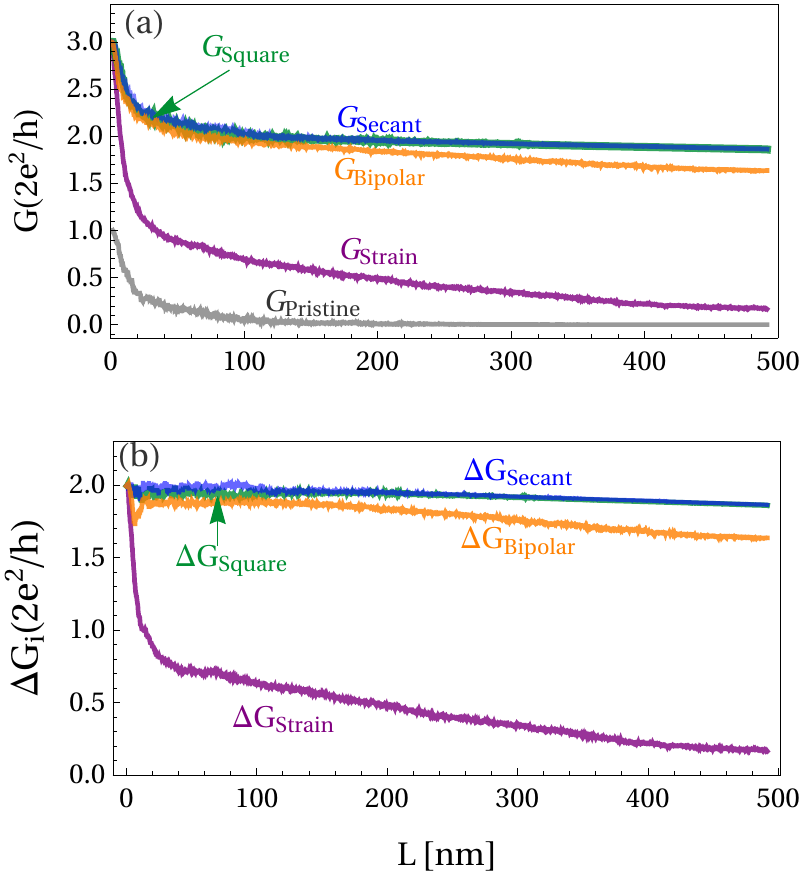}
        \caption{\label{fig:GvsL_Potential_EdgeDisoder}
         (Color) Conductance $G(\tilde E)$ and difference $\Delta G_{i}=G_{i}-G_{\text{Pristine}}$ as a function of the nanoribbon length ($x$-direction). In these panels, $G$ and $\Delta G_{i}$ are numerically obtained using edge disorder over an average of $100$ realizations. In panel (a), the solid lines corresponds the average in $G$ for: strain fold with a fixed energy $\tilde{E}=3.2\times10^{-2}$ (purple), square potential well at $\tilde{E}=3.0\times10^{-2}$ (green), hyperbolic-secant potential with $\tilde{E}=3.0\times10^{-2}$ (blue) and bipolar potential using $\tilde{E}=2.5\times10^{-2}$ (orange). The solid gray line corresponds to the behavior of pristine conductance in the presence of edge disorder. In panel (b), we show the difference $\Delta G_{i}$ between the conductance of the waveguides ($G_{i}$) and the pristine conductance ($G_{\text{Pristine}}$).}
    \end{figure}    
In this section, we study the waveguide conductance behavior $G(\tilde E)$ in the presence of edge disorder in a zGNR with dimensions $L=54.8$ nm and $W=25.8$ nm. \VI{For} comparison, conductance and band structure for a pristine zGNR is plotted in  Fig. \ref{fig:Conductance}, dotted black line exhibits the typical conductance  quantization for a zGNR, described by $G\left(\tilde{E}\right)=2n_{\tilde{E}}+1$ where $n_{\tilde{E}}\in \mathds{N}$ and $\tilde{E}=E/|t_0|$. The dotted black line, in the right panels of Fig. \ref{fig:Conductance}, displays the corresponding band structure for a pristine zGNR. Conductance and band structure in the presence of waveguiding effects are shown as solid orange lines (Fig. \ref{fig:Conductance}). We show three cases: Gaussian strain fold ($U(y_i)=0$, $A=0.7$ nm and $b=1.4$ nm) [Fig. \ref{fig:Conductance} (a) and (b)]; square potential well ($t_{ij}=t_0$, $u_1=u_2=0$ [Fig. \ref{fig:Conductance} (c) and (d)], $u_3=0.17|t_{0}|$ and $W_2=40a_0$); secant-hyperbolic potential well ($t_{ij}=t_0$, $u_2=u_3=0$, $d_1=d_2=0$, $u_1=0. 17|t_{0}|$ and $W_1=6.73a_0$) [Fig. \ref{fig:Conductance} (e) and (f)]; and bipolar potential ($t_{ij}=t_0$, $u_1=u_2=0.17|t_0|$, $u_3=0$, $d_1=d_2=26.69a_{0}$, and $W_1=6.73a_0$) [Fig. \ref{fig:Conductance} (g) and (h)]. We can observe that each waveguide modifies the energy spectrum in each valley (the first valley is centered at $ka(\pi) \approx 0.66$ and the second at $ka(\pi) \approx 1.33$) concerning the pristine case (dotted black line). Alternatively, since the conductance is proportional to the positive slope in the energy spectrum, from it we can calculate the contribution of modes associated with waveguides that modify electronic transport in zGNR.

Hence, in the case of the Gaussian strain fold [Fig. \ref{fig:Conductance}(a)], there is a decrease in the energy width of the first plateau and an increase in the width of the second plateau; this is congruent with the change shown in its energy spectrum [Fig. \ref{fig:Conductance}(b)]. As regards the single potential defined waveguides: square and secant-hyperbolic potentials, we can observe equivalent behavior in the conductance [Figs. \ref{fig:Conductance}(c)-(e)] and energy spectrum [Figs. \ref{fig:Conductance}(d)-(f)]. The latter because we have chosen widths and amplitudes values equal to each other in their definition. Therefore, in both cases, we can observe a shift to negative energies in the first plateau and an increase in the width for the second plateau of $G(\tilde E)$; this is a consequence of the presence of well-type potentials that shift the energy states to negative values. In the bipolar case, since the waveguide is defined by combination of positive and negative secant-hyperbolic potentials, we can notice a symmetric behavior of conductance concerning negative and positive energy values. However, as in the previous cases, this waveguide modifies the conductance plateaus. Thus, the first plateau decreases its width and the second plateau increases the energy width. In addition, a peculiar behavior in this waveguide is the new channels at low energies with $G(\tilde E)=5$. This conduct can be explained by two general modifications in the energy spectrum in each valley (where $G(\tilde E)=5$). First, there is a modification of the concavity in some spectrum lines, and second, there is a shift of states to low energies concerning the pristine spectrum. Therefore, these modifications in small intervals of $\tilde E$ increase the conductance in the nanoribbon.

From the above behavior of $G(\tilde E)$ under the waveguides, we can observe plateaus with $G(\tilde E)=3$, which have increased their energy width. Therefore, these plateaus have additional conductance channels due to the propagating modes of the waveguides. Thus, to estimate an energy value ($\tilde E$) associated with these extra modes (one or more per valley shown in the band structure), we perform an energy sweep to analyze the LDOS in the interval where the plateau has the value $G(\tilde E)=3$, and find the value of $\tilde E$ where the LDOS is concentrated in the center if the ribbon. Therefore, returning to Fig. \ref{fig:Conductance}, We mark these particular values of $\tilde E$ with vertical dashed lines where LDOS has its maximum behavior. These values are $\tilde E=3.2\times 10^{-2}$ for the Gaussian formation fold, $\tilde E=3.0\times 10^{-2}$ for the square well and secant-hyperbolic well potentials, and $\tilde E=2.5\times 10^{-2}$ for the bipolar potential; and thus in Fig. \ref{fig:LDOS} we show the LDOS profile across the nanoribbon (left panels) and LDOS bidimensional color map (right panels).

The behavior of $G(\tilde E)$ considering the combined effects of waveguiding and edge disorder (vacancies) is shown in the left panels of Fig. \ref{fig:Conductance} as solid green lines. Plots show the average over $100$ realizations, as described in section \ref{sec:II}. We observe that the transport properties in the nanoribbon are significantly modified. Edge disorder breaks spatial symmetry and generates local confinement suppressing  ballistic transport and therefore the quantization of the conductance channels disappears\cite{Mucciolo2009}. These effects are particular stronger in the edge states because they are localized in the edges. These effects are observed in conductance $G(\tilde E)$  for each waveguide, i.e., the conductance of the first plateau is suppressed, and in most cases, the conductance quantization is lost. However, we can observe that the plateaus with $G(\tilde E)=3$ decrease their conductance by just one unit in the presence of edge disorder and maintain their quantization in $G(\tilde E)=2$. This approximated  quantization is due to encapsulation of propagating modes by the waveguides, making these plateaus robust against edge disorder.

To evaluate the robustness of this behavior, in Fig. \ref{fig:GvsL_Potential_EdgeDisoder} we study the conductance as a function of the nanoribbon length for particular energy $\tilde{E}$ (where the LDOS is highly concentrated in the center of the ribbon). Results show a quasi-ballistic transport, in Fig. \ref{fig:GvsL_Potential_EdgeDisoder}(a) conductance in presence of waveguiding effects displays an initial value  of $G(L)=3$ and an exponential fast decay for $L<100\,$nm. This drastic drop in the conductance is due to the destruction of the channels associated with the edge states \cite{mucciolo2010disorder}. For $L>100\,$nm, the conductance decreases slowly because the waveguide reduces the effects of edge disorder and diffusion. The gray solid line shows the conductance in the absence of waveguiding effects, for this case the conductance goes fast to zero. Additional to this discussion, in  Fig. \ref{fig:GvsL_Potential_EdgeDisoder}(b) we report the difference $\Delta G_{i}=G_{i}-G_{\text{Pristine}}$. Here, $\Delta G_{i}$ can be understood as the behavior of the waveguides modes in zGNR under edge disorder. Therefore, we can observe that for the waveguides: square well, secant-hyperbolic well, and bipolar exists robust modes in $G(L)$ even when increasing the length of the nanoribbon. Finally, we can comment that the propagating waveguide modes on the center of the nanoribbon $y=0$ exhibit the most robust modes under this type of disorder.

\section{Wave guiding mechanisms bulk disorder} \label{sec:IV}
 \begin{figure}[t]
        \centering
        \includegraphics[width=.47\textwidth]{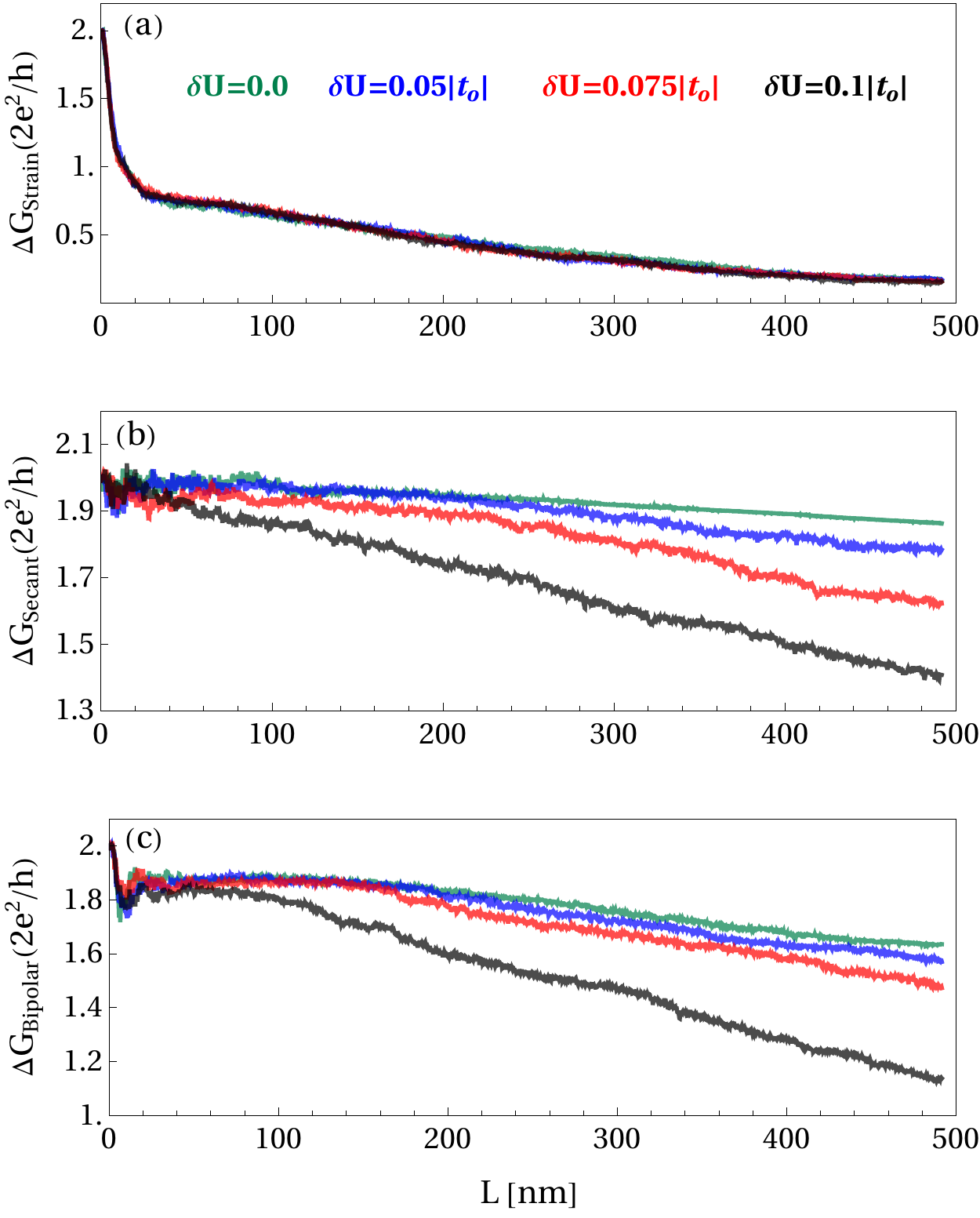}
        \caption{\label{fig:GvsL_Potential_EdgeBulkDisoder} (Color) Difference $\Delta G_{i}=G_{i}-G_{\text{Pristine}}$ as a function of the nanoribbon length ($x$-direction). In these panels, $\Delta G_{i}$ are obtained using edge and bulk disorders over an average of $100$ realizations. In panel (a), we show the analysis of $\Delta G_{\text{Strain}}$  with $\tilde{E}=3.2\times10^{-2}$, panel (b) corresponds to $\Delta G_{\text{Secant}}$ for $\tilde{E}=3.0\times10^{-2}$, and (c) in the case of $\Delta G_{\text{Bipolar}}$ with  $\tilde{E}=2.5\times10^{-2}$. In all panels, the solid blue line corresponds to the effect of the edge and bulk disorder with $\delta U=0.05|t_0|$, and the solid red line corresponds to the case $\delta U=0.075|t_0|$ and the solid black with $\delta U=0.1|t_0|$. The scatterer density in all cases is $n_{imp}=0.04$.}
\end{figure}

Aside edge defects, irregularities in the bulk can also affect the electronic properties in graphene. Bulk impurities, such as charged impurities or ripples \cite{katsnelson2008electron} can destroy the ballistic transport in graphene samples. In this section, we consider zGNRs with both, edge and bulk disorder, and analyze their electronic transport behavior in the presence of waveguiding effects: strain Gaussian fold, secant-hyperbolic potential well, and bipolar potential. Here we omit the case of the square potential well since it gives similar results to that secant-hyperbolic potential well. To introduce both disorder effects, we use the model described in section \ref{sec:II} for the cases of $\delta U=0$ (Edge disorder only), $\delta U=0.05|t_0|$, $\delta U=0.075|t_0|$, and $\delta U=0.1|t_0|$ of the random potential strength for the bulk disorder. 

In Fig. \ref{fig:GvsL_Potential_EdgeBulkDisoder} we present the behavior of the propagating modes of waveguides $\Delta G_{i}$ under edge and bulk disorder as a function of the length of the nanoribbon. We present the cases of deformation fold [Fig. \ref{fig:GvsL_Potential_EdgeBulkDisoder} (a)], secant-hyperbolic potential [Fig. \ref{fig:GvsL_Potential_EdgeBulkDisoder} (b)] and bipolar potential [Fig. \ref{fig:GvsL_Potential_EdgeBulkDisoder} (c)]. Thus, for the strain fold, we can observe that the values of $\delta U$ for bulk disorder do not affect the response of $\Delta G_{\text{Strain}}$ compared to edge disorder ($\delta U=0$). This robust behavior is possible due to the inhomogeneous pseudomagnetic field\cite{carrillo2016strained} generated by the Gaussian strain in the zGNR, that bends the electron trajectories. However, for $\delta U>0.26|t_0|$ (not shown), the response of $\Delta G_{\text{Strain}}$ as a function of $L$ decays faster than the curves shown in Fig. \ref{fig:GvsL_Potential_EdgeBulkDisoder}(a). Nevertheless, in the case of strain fold and just bulk disorder, increasing $\Delta U$ decreases $\Delta G_{\text{Strain}}$. On the other hand, due to the nature of the secant-hyperbolic well and bipolar potentials, which can be understood as electrostatic fields in the zGNR; they do not generate a pseudomagnetic field and the effect of bulk disorder decreases $\Delta G_{\text{Secant}}$ and $\Delta G_{\text{Bipolar}}$ as a function of nanoribbon length. However, the response of the bipolar potential is more sensitive to edge and bulk disorder with respect to the case of the secant-hyperbolic well because the mode is more spread withing the nanoribbon.

\section{Conclusions}\label{sec:V}

In summary, we have studied electronic transport in zGNRs under the effect of waveguides and the presence of edge and bulk disorder. We review the cases of the strain fold, square well,  hyperbolic-secant well, and bipolar potentials. We notice that for $\tilde{E}>0$; the number of ballistic channels increases due to propagating modes in these waveguides. This extra channels are modified edge states by the waveguide. In the presence of edge and bulk disorder, the conductance decreases; nevertheless, there are propagating modes associated with the waveguide, exhibiting quasi-ballistic transport. In particular, the square well, hyperbolic-secant well, and bipolar potentials have more robust protected modes, even increasing the nanoribbon length. We hope our work could lead to improve graphene-based electronic devices. \cite{williams2011gate}.

\section{Acknowledgements}\label{sec:Acknowledgements}

E.J.R.-R acknowledges financial support from CONACyT.
R.C.-B. thank 20va Convocatoria
Interna (UABC) 400/1/64/20. V.G.I.-S and J.C.S.-S. acknowledge the total support from Estancias Posdoctorales por M\'exico 2021 CONACYT.

\bibliography{referencias.bib}

\end{document}